\documentclass[aps,groupedaddress]{revtex4}
\usepackage{epsfig}
\usepackage{graphicx}
\usepackage[T1]{fontenc}
\usepackage{ae}
\usepackage{color}
\usepackage[latin1]{inputenc}
\usepackage{amssymb,amsbsy,amsmath}
\usepackage{bbm}
\usepackage{undertilde}

\newtheorem{lemma}{Lemma}

\newtheorem{prop}{Proposition}

\newcommand{\kB}{k_{\rm B}}
\begin{document}

%\begin{center}\today\end{center}

%%%%%%%%%%%%%%%%%%%%%%%%%%%%%%%%%%%%%%%%%%%%%%%%%%%%%%%%%%%%%%%%%%%%%%%%%%%%%

\title[Quantum Maxwell's demon]
{Conditional action and quantum versions of Maxwell's demon}

\author{Heinz-J\"urgen Schmidt
}
\address{ Universit\"at Osnabr\"uck,
Fachbereich Physik,
 D - 49069 Osnabr\"uck, Germany
}

%\tableofcontents

\begin{abstract}
We propose a new way of looking at the quantum Maxwell's demon problem in terms of conditional action.
A ``conditional action" on a system is a unitary time evolution,
selected according to the result of a previous measurement,
which can reduce the entropy of the system. However, any conditional action
can be realized by an (unconditional) unitary time evolution
of a larger system and a subsequent L\"uders measurement,
whereby the entropy of the entire system is either increased or remains constant.
We give some examples that illustrate and confirm our proposal,
including the erasure of $N$ qubits and the Szilard engine, 
thus relating the present approach to the Szilard principle and the Landauer principle
that have been discussed as possible solutions of the Maxwell's demon problem.
\end{abstract}

\maketitle

%%%%%%%%%%%%%%%%%%%%%%%%%%%%%%%%%%%%%%%%%%%%%%%%%%%%%%%%%%%%%%%%%%%%%%%%%%%%%%%%%%%%%%%%%%%%%%%%%%%%%%%%%%%%%%%%%%%%%%%%%%%%%%
\section{Introduction}\label{sec:I}
%%%%%%%%%%%%%%%%%%%%%%%%%%%%%%%%%%%%%%%%%%%%%%%%%%%%%%%%%%%%%%%%%%%%%%%%%%%%%%%%%%%%%%%%%%%%%%%%%%%%%%%%%%%%%%%%%%%%%%%%%%%%%%%

Since its first appearance in 1867, the thought experiment of James Clerk Maxwell has given rise to many ideas and probably more than a thousand papers
\cite{footnote}.
In the thought experiment a demon controls a small door between two gas chambers. When single gas molecules approach the door,
the demon opens and closes the door quickly, so that only fast molecules enter one of the chambers, while slow molecules enter the other one.
In this way the demon's behavior causes one chamber to heat up and the other to cool down,
reducing entropy and violating the  $2^{nd}$ law of thermodynamics.

\begin{figure}[t]
\centering
\includegraphics[width=0.7\linewidth]{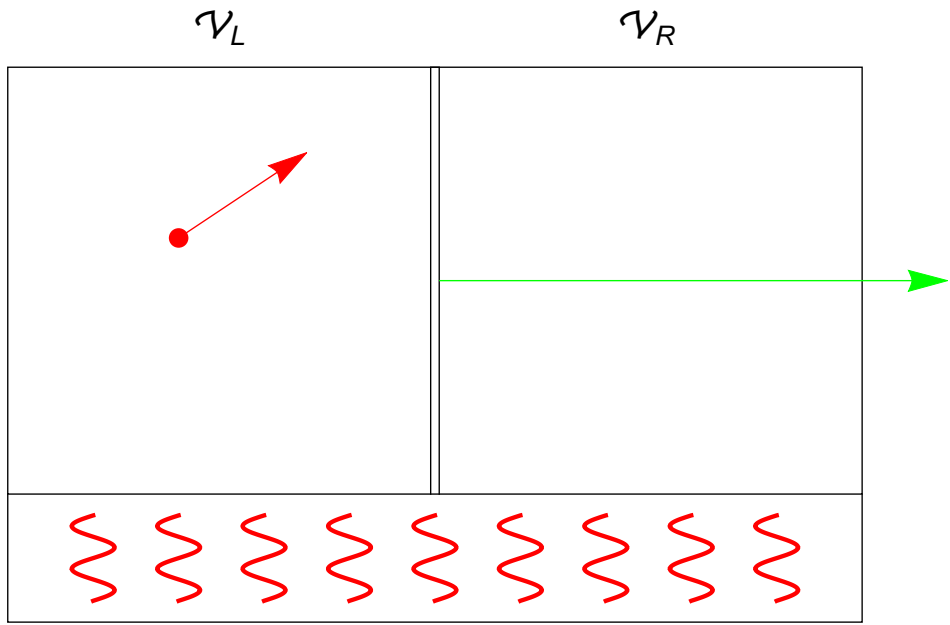}
\caption{Schematic representation of Szilard's engine. A volume is separated by a piston into two chambers ${\mathcal V}_R$
and  ${\mathcal V}_L$ of equal volume. A molecule is localized by a measurement in one of these chambers and, depending on the result of the measurement,
an isothermal expansion of the one-molecule gas in contact with a heat bath will be performed to the right or to the left. In the figure we show the case where
the molecule has been found in the left chamber and the expansion is performed to the right.
}
\label{FIGSZ}
\end{figure}

Among the most influential defenses of the $2^{nd}$ law are those of Szilard \cite{S29} and Landauer/Bennett \cite{L61},\cite{B82}.
Szilard proposes his own version (``Szilard's engine") of the original thought experiment that consists only of one gas particle which can be found in
the right or the left chamber of a cylindrical box divided by a piston. Depending on its position an isothermal expansion of the
one-molecule gas is performed to the left or to the right thereby converting heat from a heat bath completely into work, see Figure \ref{FIGSZ}.
Szilard argues that the entropy decrease of the system is compensated by the entropy costs of acquiring information about the position
of the gas particle (``Szilard's principle"). His arguments are formulated within classical physics and not easy to understand, see
also the analysis and reconstruction of Szilard's reasoning in \cite{LR94} and \cite{EN99}.

Based on Landauer's calculations  \cite{L61} on the thermodynamics of computing
Bennett has shifted the focus from the entropy costs of acquiring to {\em erasing} information \cite{B82}.
He argues that for a cyclic operation of a Szilard engine converting heat completely into work the memory device that
contains the information about the initial measurement should be set to a default value each time. This erasure
of information produces at least the entropy needed to compensate the entropy decrease caused by the engine.
This explanation (``Landauer's principle") has today been adopted by the main stream of physicists, but has also been
criticized by a minority of scholars, see \cite{EN98}, \cite{EN99},  \cite{N19} and further references cited there.
For the present paper it will be sensible to distinguish between the principle that erasure of memory produces entropy
(``Landauer's principle" in the narrow sense) and the position that this effect constitutes the solution of the apparent
paradox of Maxwell's demon (henceforward called ``Landauer/Bennett principle").

Whereas the arguments of Szilard and  Landauer/Bennett are mainly classical,
it appears plausible that a proper account of entropy increase due to measurements should be discussed within the realm of quantum theory.
A first attempt of a quantum-theoretical account of Szilard's engine has been given by W.~H.~Zurek \cite{Z84},
followed by \cite{L87} - \cite{L97}. More recently, the paradigm of Maxwell's demon has been used in connection with
quantum information theory, especially quantum error correction, see \cite{V00} and \cite{NC00}.

Zurek in his  \cite{Z84} considered
a one-particle quantum system in a box described by a Gibbs ensemble and  calculated the increase
of free energy due to the measurement of whether the particle is in the right or in the left chamber.
In the section of his paper headlined
``Measurement by `quantum Maxwell's demon' " Zurek presented a model of the measurement using ideas of decoherence
and finally also incorporated the Landauer/Bennett issue of memory erasure.
However, the complete entropy balance remains opaque.
In terms of content, it would be plausible to regard the paper as a quantum mechanical
justification of the Szilard principle. But then the statement in the summary
%\newpage
\begin{quotation}\small
  ``Moreover, we show that the ultimate reason for
the entropy increase can be traced back to the necessity to `reset' the state
of the measuring apparatus, which, in turn, must involve a measurement."
\end{quotation}
\normalsize
would appear as an unfounded tribute to the Landauer/Bennett principle.
Therefore the general message is not quite clear.
Further, there are three questions left open:
\begin{itemize}
  \item Are the information-theoretic concepts used in \cite{Z84} only an illustration of the theoretical account or are they crucial to solve the
  Maxwell's demon problem? This question is the more important since there exist suggestions of extending the framework of statistical mechanics
  by information-theoretic notions, see, e.~g., \cite{SU10},\cite{SS15}.
  \item Similarly, are the ideas from theories of decoherence, see also \cite{L87} and \cite{L89},  really necessary to solve the  Maxwell's demon problem?
  \item Since the paper follows very closely the details of Szilard's engine, one wonders which assumptions and approximations
   are decisive for the solution presented and which are only made for convenience.
   In other words, a more abstract representation of the ``quantum Maxwell's demon'' would be desirable.
\end{itemize}

In the present paper we will pursue a similar approach but try to amend and extend Zurek's results in the way indicated above.
Our explanation of the apparent paradoxical results of Maxwell's demon acting on a quantum system (also called ``object system")
will be given in three steps:
\begin{itemize}
  \item First we define the concept of ``conditional action" that comprises the original version of Maxwell's demon as well as
  Szilard's engine and Landauer's erasure of memory. The mathematical representation of ``conditional action" on quantum systems
  results in a special kind of {\em instruments}, in the sense of \cite{BLPY16},  that we will call ``Maxwell instruments".
  \item We show that the total operation of a Maxwell instrument {\em may} decrease the von Neumann entropy of the object system
  depending on the initial state. If this happens we will call the Maxwell instrument ``demonic".
  \item A demonic Maxwell instrument always has a physical realization of the following kind: The object system
  is extended by an auxiliary system and the total system undergoes a unitary time evolution followed by a L\"uders measurement
  at the auxiliary system. If reduced to the object system the final state will have a smaller entropy than at the beginning
  although the total entropy will increase in accordance with what a $2^{nd}$ law of quantum mechanics presumably would predict.
\end{itemize}

It has been criticized \cite{EN98},\cite{EN99}  that the Landauer/\-Bennett defense of the $2^{nd}$ law
against Maxwell's demon in turn presupposes the $2^{nd}$ law.
We avoid these pitfalls of circularity since we do not assume any general $2^{nd}$ law
in quantum mechanics but only a few well-established
theorems about the increase of von Neumann entropy during L\"uders measurements and state separation.
Actually we would not know how to formulate such a general $2^{nd}$ quantum law.
In this respect the role of Maxwell's thought experiment is different in classical and in quantum theory:
In classical theory it is a potential paradox since it seems to contradict the well-established $2^{nd}$ law.
In quantum theory it is rather a tool to find such a general $2^{nd}$ law.
Fortunately, Maxwell-demonic interventions can be
formalized within the realm of quantum measurement theory where already fragments of a $2{nd}$ law exist
that are sufficient to explain the demon's actions.

The paper is organized as follows: In Section \ref{sec:OI} we recapitulate some well-known definitions and results
from quantum measurement theory for the convenience of the reader. These concepts are applied in Section \ref{sec:QMD}
to explain why the conditional action of Maxwell' demon possibly lowers the entropy of the object system but leads to an at least
equal amount of entropy increase in some auxiliary system. The following section \ref{sec:EX} contains two simple examples
illustrating the former considerations. A classical version of ``conditional action" will be sketched in Section \ref{sec:C},
followed by a Summary in Section \ref{sec:SU}. We have deferred some proofs (\ref{sec:M},\ref{sec:P}) and the explicit construction of a measurement
dilation of a Maxwell instrument (\ref{sec:E}) into the Appendix, as well as the detailed account of Szilard's engine (\ref{sec:S})
according to our approach.

%%%%%%%%%%%%%%%%%%%%%%%%%%%%%%%%%%%%%%%%%%%%%%%%%%%%%%%%%%%%%%%%%%%%%%%%%%%%%%%%%%%%%%%%%%%%%%%%%%%%%%%%%%%%%%%%%%%%%%%%%%%%%%
\section{Operations and instruments}\label{sec:OI}
%%%%%%%%%%%%%%%%%%%%%%%%%%%%%%%%%%%%%%%%%%%%%%%%%%%%%%%%%%%%%%%%%%%%%%%%%%%%%%%%%%%%%%%%%%%%%%%%%%%%%%%%%%%%%%%%%%%%%%%%%%%%%%%

In the following sections we will heavily rely upon the mathematical notions of {\em operations} and {\em instruments}.
Although these notions are well-known it will be in order to recall the pertinent definitions adapted to the present purposes
and their interpretations in the context of measurement theory. In order to keep the presentation as simple as possible we restrict ourselves to the
case of finite dimensional Hilbert spaces ${\mathcal H}$ and refer the reader to the literature on the general case of separable Hilbert spaces.

Let $B({\mathcal H})$ denote the space of Hermitean operators  $A:{\mathcal H}\longrightarrow {\mathcal H}$
and $B_+({\mathcal H})$ the cone of positively semi-definite operators, i.~e., having only non-negatives eigenvalues. Moreover, let
$T:B({\mathcal H})\longrightarrow B({\mathcal H})$ be a linear map satisfying
\begin{itemize}
  \item $T$ is {\em positive}, i.~e., maps  $B_+({\mathcal H})$ into itself,
  \item $T$ is {\em completely positive}. This means that
  $T\otimes {\mathbbm 1}:B({\mathcal H}\otimes {\mathbbm C}^N)\longrightarrow B({\mathcal H}\otimes {\mathbbm C}^N)$
  will be positive for all integers $N$.
\end{itemize}
Then $T$ will be called an {\em operation}. It may be trace-preserving or not.

Operations are intended to describe state changes due to measurements. By definition, a {\em L\"uders measurement}
(without selection according to the outcomes) induces the state change
\begin{equation}\label{OI1}
  \rho \mapsto L(\rho)=\sum_{n\in{\mathcal N}} P_n\,\rho\,P_n
  \;,
\end{equation}
where $\left(P_n\right)_{n\in{\mathcal N}}$ denotes a complete family of mutually orthogonal projections $P_n\in B_+({\mathcal H})$.
The L\"uders operation $L$ is an example of a trace-preserving operation. Note that the map (\ref{OI1}) is defined for all
$\rho\in B({\mathcal H})$ whereas the physical interpretation holds only for {\em statistical operators} $\rho$,
i.~e., for positively semi-definite operators with $\mbox{Tr}(\rho)=1$.

We mention the following representation theorem for operations, see, e.~g., \cite{BLPY16}, prop.7.7, or
\cite{NC00}, chapter 8.2.3. $A$ is an operation iff it can be written as
\begin{equation}\label{OI2}
 A( \rho)= \sum_{i\in{\mathcal I}}A_i\,\rho\,A_i^\ast
  \;,
\end{equation}
with the {\em Kraus operators} $A_i:{\mathcal H}\rightarrow{\mathcal H}$ and a finite
index set ${\mathcal I}$. Comparison of (\ref{OI1}) and (\ref{OI2}) shows that for the L\"uders operation one may choose
${\mathcal I}={\mathcal N}$ and $A_n=P_n$ for all $n\in{\mathcal N}$.

In (\ref{OI1}) we have considered the total state change without any selection. If we select according to
the outcome of the L\"uders measurement we would obtain a family of (not trace preserving) operations
\begin{equation}\label{OI3}
  L_n(\rho)= P_n\,\rho\,P_n,\quad n\in{\mathcal N}
\;,
\end{equation}
that describe conditional state changes. This situation can be generalized in the following way.

Let ${\mathcal N}$ be a finite index set. Then the map
${\mathfrak I}:{\mathcal N}\times B({\mathcal H})\longrightarrow B({\mathcal H})$
will be called an {\em instrument} iff
\begin{itemize}
  \item ${\mathfrak I}(n)$ is an operation for all $n\in{\mathcal N}$, and
  \item $\mbox{Tr}\left(\sum_{n\in{\mathcal N}}{\mathfrak I}(n)(\rho)\right)=\mbox{Tr}\rho$ for all $\rho\in B({\mathcal H})$.
\end{itemize}
The second condition can be rephrased by saying that the {\em total operation} ${\mathfrak I}({\mathcal N})$ defined by
\begin{equation}\label{OI4}
  {\mathfrak I}({\mathcal N})(\rho)\equiv \sum_{n\in{\mathcal N}}{\mathfrak I}(n)(\rho)
\end{equation}
will be trace-preserving. The special case (\ref{OI3}) will be referred to as a {\em L\"uders instrument}.

The comparison with the definition 7.5 of \cite{BLPY16} shows that, besides neglecting
convergence conditions, we have specialized the general definition of an instrument to the case of
a finite outcome space ${\mathcal N}$. Measurements of continuous observables
like position or momentum would require to consider elements of the $\sigma$-algebra of Borel subsets of, say, ${\mathbbm R}^N$
for the first argument of the instrument. This generalization is not necessary to be considered in the present paper.

We will need a second representation theorem, this time formulated for instruments.
It is called a {\em measurement dilation} and can be physically viewed as a realization of a
non-L\"uders instrument ${\mathfrak J}$ by a time evolution and a L\"uders instrument on a larger system. Thus let ${\mathcal K}$
be another Hilbert space, $\phi\in {\mathcal K}$ a vector with $\|\phi\|=1$ and corresponding projection $P_\phi$ and
$V:{\mathcal H}\otimes {\mathcal K}\longrightarrow {\mathcal H}\otimes {\mathcal K}$ a unitary operator.
Further, let $\left(Q_n\right)_{n\in{\mathcal N}}$ be a complete family of mutually orthogonal
projections in ${\mathcal K}$. Then the  map  ${\mathfrak D}_{{\mathcal K},\phi,V,Q}:{\mathcal N}\times B({\mathcal H})\longrightarrow B({\mathcal H})$
defined by
\begin{eqnarray}\label{OI5}
  && {\mathfrak D}_{{\mathcal K},\phi,V,Q}(n)(\rho)\\
  &\equiv& \mbox{Tr}_{\mathcal K}\left(
  \left( {\mathbbm 1}\otimes Q_n\right)V\left(\rho\otimes P_\phi \right)V^\ast\left( {\mathbbm 1}\otimes Q_n\right)
  \right)
\end{eqnarray}
will be an instrument. Here $\mbox{Tr}_{\mathcal K}$ denotes the {\em partial trace} that reduces a state of the total system
to a state of the subsystem given by the Hilbert space ${\mathcal H}$. If ${\mathfrak J}$ is a given instrument then
${ \mathfrak D}_{{\mathcal K},\phi,V,Q}$ will be called a {\em  measurement dilation} of ${\mathfrak J}$ iff
${\mathfrak J}={\mathfrak D}_{{\mathcal K},\phi,V,Q}$. The mentioned representation theorem
guarantees the existence of measurement dilations for any given instrument, see Theorem 7.~14 of  \cite{BLPY16}
or Exercise 8.~9 of \cite{NC00}. The last reference also contains an explicit construction procedure for
${ \mathfrak D}_{{\mathcal K},\phi,V,Q}$ that will be reproduced for the special case of a Maxwell instrument
in Appendix \ref{sec:E} and will henceforward be referred to as the ``standard realization".

%%%%%%%%%%%%%%%%%%%%%%%%%%%%%%%%%%%%%%%%%%%%%%%%%%%%%%%%%%%%%%%%%%%%%%%%%%%%%%%%%%%%%%%%%%%%%%%%%%%%%%%%%%%%%%%%%%%%%%%%%%%%%%
\section{The quantum version of Maxwell's demon (QMD)}\label{sec:QMD}
%%%%%%%%%%%%%%%%%%%%%%%%%%%%%%%%%%%%%%%%%%%%%%%%%%%%%%%%%%%%%%%%%%%%%%%%%%%%%%%%%%%%%%%%%%%%%%%%%%%%%%%%%%%%%%%%%%%%%%%%%%%%%%%

The activity of Maxwell's demon can be abstractly characterized as performing a {\em conditional action}, i.~e., an action depending on the results
of a previous measurement. Additionally, it is required that this conditional action leads to an entropy decrease of the system
if applied to a certain set ${\mathcal A}$ of {\em admissible initial states}. In this paper we will interpret these notions quantum mechanically,
especially the states as statistical operators $\rho$ of a so-called object system defined
on some Hilbert space ${\mathcal H}$, and the measurement as a L\"uders instrument
\begin{equation}\label{QMD1}
  \mathfrak{I}(n)(\rho)={P}_n\,\rho\,{P}_n
  \;,
\end{equation}
where $n$ runs through some finite index set ${\mathcal N}$ and
$\left({P}_n\right)_{n\in{\mathcal N}}$ is a complete family of mutually orthogonal projections.
The total L\"uders operation
\begin{equation}\label{QMD2}
  \mathfrak{I}({\mathcal N})(\rho)=\sum_{n\in{\mathcal N}}{P}_n\,\rho\,{P}_n
  \end{equation}
represents the state change after the L\"uders measurement without any selection.
More general instruments may be used to model the demon's measurement but this
possibility  will not be considered in the present paper.

Further, the entropy is taken as the von Neumann entropy \cite{vN32}
\begin{equation}\label{QMD3}
  S(\rho)=-{\mbox Tr}\left(\rho\,\log\,\rho\right)
  \;,
\end{equation}
where $\log$ is chosen as the natural logarithm.
It is well-known \cite{vN32}, \cite{NC00}, \cite{SG20} that the entropy of a state never decreases during a L\"uders measurement, i.~e.,
\begin{equation}\label{QMD4}
 S(\rho)\le S\left(\sum_{n\in{\mathcal N}}{P}_n\,\rho\,{P}_n \right)\equiv \widetilde{S}_1
 \;.
\end{equation}
Hence a L\"uders measurement alone cannot be used to model a QMD. Additionally,
we need to give a quantum-theoretical definition of a {\em conditional action} relative to a L\"uders measurement.
This will be done by considering a family $\left( U_n\right)_{n\in{\mathcal N}}$ of unitary
operators in ${\mathcal H}$ such that the combined state change will be given by the instrument
\begin{equation}\label{QMD5}
 \mathfrak{J}(n)(\rho)= U_n\,P_n\,\rho P_n\,U_n^\ast
 \;,
\end{equation}
henceforward called a ``Maxwell instrument",
with total operation (``Maxwell operation")
\begin{equation}\label{QMD6}
 \rho\mapsto\rho_1=\mathfrak{J}({\mathcal N})(\rho)=\sum_{n\in{\mathcal N}} U_n\,P_n\,\rho P_n\,U_n^\ast
 \;.
\end{equation}
Again the Kraus operators $A_n=U_n\,P_n$ of the operation $\mathfrak{J}({\mathcal N})$ may be read off the representation (\ref{QMD6}).

We stress that we will use the mathematical notion of an instrument that was originally designed to characterize
state changes due to measurements in order to describe the more general state changes caused by a measurement {\em and} a conditional action.
A similar approach has been adopted in chapter 12.4.4 of \cite{NC00} in connection with quantum error correction.

It can be shown that a Maxwell operation always decreases the entropy of the corresponding post-measurement state:
\begin{prop}\label{propMAX}
\begin{equation}\label{MAX1}
 S_1\equiv S\left(\sum_{n\in{\mathcal N}} U_n\,P_n\,\rho P_n\,U_n^\ast\right)\le S\left(\sum_{n\in{\mathcal N}}P_n\,\rho P_n\right)=\widetilde{S}_1
 \;.
\end{equation}
\end{prop}
For a proof see Appendix \ref{sec:P}.

It is obvious that the $U_n$ are not uniquely determined by
(\ref{QMD5}), for example, $U_n$ must only be defined on the support of $P_n$ and can be arbitrarily extended
to its orthogonal complement.
In other words: the conditional action must be only defined for those cases where the condition holds.

In passing we note that the concept of ``conditional action" is also used in quantum teleportation, see \cite{NC00}, chapter 1.3.7.
Here Alice makes two quantum measurements and sends her results to Bob via a classical communication channel,
who in turn performs certain operations depending on the measurement results. However,
the total entropy increases during teleportation and hence it cannot be considered as a QMD.

It is well-known that in the case of a  more general instrument than that of L\"uders type  a statement analogous to
(\ref{QMD4}) may fail, i.~e., a generalized measurement can decrease entropy, see \cite{NC00}, Exercise 11.15.
We will provide two examples in Section \ref{sec:EX} showing that this may also happen for an instrument of the
form (\ref{QMD6}) and hence the Maxwell instrument is a possible candidate for a QMD.

We know from classical thermodynamics that the decrease of entropy of some system would not contradict the $2^{nd}$ law of thermodynamics
if it is accompanied by an equal or larger increase of entropy in some other parts of the world.
This strategy of explaining the decrease of entropy can also be tried in the case of quantum mechanics.
It is highly plausible that the demon needs some auxiliary system to perform the measurement
and the conditional action. We will call this auxiliary system again the ``demon" and  assume that it can
be modelled as another quantum system with Hilbert space ${\mathcal K}$. How can the quantum demon be realized?
It is tempting to use the {\em measurement dilation} sketched in Section \ref{sec:OI} that was originally
intended to merely  give a physical realization of a non-L\"uders measurement.
But there is no reason not to apply this construction to Maxwell instruments ${\mathfrak J}$ as well.

Hence we will assume that at the beginning the state of the combined system, object system and demon, is assumed to be
\begin{equation}\label{QMD8}
  \rho\otimes P_\phi
  \;,
\end{equation}
where $P_\phi$ is a one-dimensional projector in ${\mathcal K}$. Then a unitary time evolution $V$ of the combined system
takes place with the resulting state being
\begin{equation}\label{QMD9}
  V\,\left(  \rho\otimes P_\phi\right)\,V^\ast
  \;,
\end{equation}
followed by a L\"uders measurement at the demon with projectors $Q_n:{\mathcal K}\rightarrow {\mathcal K}$.
This leads to a (not normalized) state
\begin{equation}\label{QMD10}
\left( {\mathbbm 1}\otimes Q_n\right)\, V\,\left(  \rho\otimes P_\phi\right)\,V^\ast\,\left( {\mathbbm 1}\otimes Q_n\right)
 \;.
\end{equation}
Finally this state is reduced to the object system by performing the partial trace $\mbox{Tr}_{\mathcal K}$.
This yields the measurement dilation of ${\mathfrak J}$ of the form
\begin{eqnarray}\nonumber
&&
 {\mathfrak D}_{{\mathcal K},\phi,V,Q}(n)(\rho)\\
 \label{QMD11}
 &\equiv&\mbox{Tr}_{\mathcal K}\left( \left( {\mathbbm 1}\otimes Q_n\right)\, V\,\left(  \rho\otimes P_\phi\right)\,V^\ast\,\left( {\mathbbm 1}\otimes Q_n\right)\right)
 \;,
\end{eqnarray}
with corresponding total operation
\begin{eqnarray}\nonumber
&&
 {\mathfrak D}_{{\mathcal K},\phi,V,Q}({\mathcal N})(\rho)\\
 \label{QMD12}
 &=&\sum_{n\in{\mathcal N}}\mbox{Tr}_{\mathcal K}\left( \left( {\mathbbm 1}\otimes Q_n\right)\, V\,
 \left(  \rho\otimes P_\phi\right)\,V^\ast\,\left( {\mathbbm 1}\otimes Q_n\right)\right)
\end{eqnarray}

Before entering into the proposed solution of the mentioned paradox we would like to point out that the measurement dilation
(\ref{QMD11}) in a sense reverses the temporal order of measurement and (conditional) action. In the original description
of the demon we imagine a measurement followed by an action depending on the result of that measurement. In the dilation
(\ref{QMD11}) there is first an unconditioned time evolution of the combined system followed by a state change due to a
L\"uders measurement at the demon and the state reduction. This resembles the difference between a classical computer
that executes an ``if-else" command thereby performing a conditional action and a quantum computer that performs all possible actions
simultaneously until a final measurement selects which condition is satisfied. Such a realization seems strange at first sight
but is a consequence of our decision to describe the demon purely as a quantum system.

Coming back to the apparent violation of a tentative $2^{nd}$ law  it is clear that the entropy of the quantum state
remains constant during the first steps of the operation ${\mathfrak D}(\mathcal N)$:
\begin{equation}\label{QMD13}
  S_0\equiv S(\rho)=S\left(\rho \otimes P_\phi\right)=S\left(V\,\left(\rho \otimes P_\phi\right)\,V^\ast\right)
  \;,
\end{equation}
since the entropy is additive for tensor products, vanishes for pure states and is unitarily invariant.
By the following L\"uders measurement the entropy increases (or remains constant)
according to (\ref{QMD4}):
\begin{eqnarray}
\label{QMD14a}
  S(\rho) &\le& S(\rho_{12})\;, \\
  \label{QMD14b}
  \rho_{12} &\equiv&
  \sum_{n\in{\mathcal N}}
  \left({\mathbbm 1}\otimes Q_n\right)\, V\,\left(  \rho\otimes P_\phi\right)\,V^\ast\,\left({\mathbbm 1}\otimes Q_n\right)
  \;.
\end{eqnarray}
If we reduce $\rho_{12}$ to both subsystems,
\begin{equation}\label{QMD15}
  \rho_{12}\mapsto \rho_1\otimes \rho_2\equiv \left(\mbox{Tr}_{\mathcal K}\rho_{12}\right)\otimes \left(\mbox{Tr}_{\mathcal H}\rho_{12}\right)
  \;,
\end{equation}
the entropy further increases:
\begin{equation}\label{QMD16}
  S_0\le S(\rho_{12})\le S(\rho_1)+S(\rho_2)
  \;.
\end{equation}
This is a consequence of the so-called subadditivity of the von Neumann entropy, see \cite{NC00}, 11.3.4.
The inequality (\ref{QMD16}) is compatible with the condition for a QMD
\begin{equation}\label{QMD17}
  S(\rho_1)<S_0
  \;,
\end{equation}
since it only implies
\begin{equation}\label{QMD18}
 S(\rho_2)\stackrel{(\ref{QMD16})}{\ge}S_0-S(\rho_1)\stackrel{(\ref{QMD17})}{>}0
 \;.
\end{equation}
This means that the decrease of the entropy of the object system will be, at least, compensated by an
increase of the demon's entropy. In this case the total entropy of the object system
and the demon does not decrease in accordance with a tentative $2^{nd}$ law.

%%%%%%%%%%%%%%%%%%%%%%%%%%%%%%%%%%%%%%%%%%
\section{Examples}
\label{sec:EX}
%%%%%%%%%%%%%%%%%%%%%%%%%%%%%%%%%%%%%%%%%%

%%%%%%%%%%%%%%%%%%%%%%%%%%%%%%%%%%%%%%%%%%
\subsection{Erasure of $N$ qubits}
\label{sec:ER}
%%%%%%%%%%%%%%%%%%%%%%%%%%%%%%%%%%%%%%%%%%
As a first example of a demonic Maxwell instrument ${\mathfrak E}$ and its standard realization we consider
a system with a Hilbert space being an $N$-fold tensor product of two-dimensional ones
\begin{equation}\label{ER1}
  {\mathcal H}=\bigotimes_{\nu=1}^N {\mathbbm C}^2_{(\nu)}
\end{equation}
and an orthonormal basis of vectors $|n\rangle,\; n\in{\mathcal N}\equiv \{0,\ldots,2^N-1\}$ where $n$ is identified with the string
of length $N$ consisting of its binary digits. Especially,  $0$ represents the string consisting of $N$ zeroes.
Further we choose an initial L\"uders measurement with projectors
\begin{equation}\label{ER2}
  P_n=|n\rangle\langle n|,\; n\in{\mathcal N}
  \;,
\end{equation}
and the unitaries $U_n$ corresponding to the Maxwell instrument (\ref{QMD5}) such that
\begin{equation}\label{ER3}
  U_n\,|n\rangle = |0\rangle
\end{equation}
for all $n\in{\mathcal N}$. After a short calculation we obtain
\begin{equation}\label{ER4}
 {\mathfrak E}({\mathcal N})(\rho) =\sum_{n\in{\mathcal N}}U_n\,P_n\,\rho\,P_n\,U_n=P_0
 \;,
\end{equation}
for all statistical operators $\rho$ in ${\mathcal H}$
and hence the description of the Maxwell instrument ${\mathfrak E}$  as ``erasure of $N$ qubits" seems adequate.
Since
\begin{equation}\label{ER5}
  S\left( {\mathfrak E}({\mathcal N})(\rho)\right) =S(P_0)=0
  \;,
\end{equation}
the entropy decrease of the corresponding Maxwell operation is maximal and we may call it ``demonic".

Its standard realization is given by ${\mathcal K}={\mathcal H}$, $\phi=|0\rangle$, $Q_n=P_n$ for all $n\in{\mathcal N}$
and
\begin{equation}\label{ER6}
  V:{\mathcal H}\otimes{\mathcal K}\longrightarrow {\mathcal H}\otimes{\mathcal K},\quad
  V\left(\phi_1\otimes \phi_2\right)= \phi_2\otimes \phi_1
  \;.
\end{equation}
After a short calculation  we obtain, in accordance with (\ref{ER4}),
\begin{equation}\label{ER7a}
\rho_1=\mbox{Tr}_{\mathcal K}\left(\rho_{12} \right)=P_0
\;,
\end{equation}
where
\begin{equation}\label{ER7b}
 \rho_{12}\equiv
  \sum_{n\in{\mathcal N}}\left({\mathbbm 1}\otimes Q_n \right)\,V\,\left( \rho\otimes P_0\right)\,V^\ast\,\left({\mathbbm 1}\otimes Q_n \right)
  \;,
\end{equation}
and
\begin{equation}\label{ER8}
 \rho_2=\mbox{Tr}_{\mathcal H}\left(\rho_{12} \right)= \sum_{n\in{\mathcal N}}P_n\,\rho\,P_n
 \;.
\end{equation}
Moreover,
\begin{equation}\label{ER9}
 S\left(\rho_2\right)= S\left(\sum_{n\in{\mathcal N}}P_n\,\rho\,P_n \right) \ge S(\rho)
 \;,
\end{equation}
by virtue of (\ref{QMD4}).

This means that the standard realization of the Maxwell instrument ${\mathfrak E}$
erasing $N$ qubits proceeds by shifting the post-measurement state of the L\"uders measurement corresponding to (\ref{ER2})
into an auxiliary system of the same size as the object system. According to (\ref{ER9}) this overcompensates
the decrease of entropy due to the erasure. Since we have not precisely stated a quantum version of Landauer's principle (in the narrow sense)
we cannot claim that this would represent a proof of this principle. A possible obstacle would be that such a principle
is usually formulated to make a statement about {\em all} possible realizations of the erasure process, whereas we have only said what would
be obtained for realizations by measurement dilations ${\mathfrak E}={\mathfrak D}_{{\mathcal K},\phi,V,Q}$.

Note finally that the usual statement about the entropic costs for erasure of at least $ \kB\, \log 2$ per bit
(re-introducing the Boltzmann constant $\kB$) follows from
\begin{equation}\label{ER10}
 S\left(\sum_{n\in{\mathcal N}}P_n\,\rho\,P_n \right)=-\sum_{n\in{\mathcal N}}p_n\,\log\, p_n
 \;,
\end{equation}
if all $p_n\equiv \mbox{Tr} \left( \rho\,P_n\right)$ are equal and hence $p_n=2^{-N}$ which entails
$-\sum_{n\in{\mathcal N}}p_n\,\log\, p_n= N\, \log 2$.

%%%%%%%%%%%%%%%%%%%%%%%%%%%%%%%%%%%%%%%%%%
\subsection{A simple model of a QMD}
\label{sec:EM}
%%%%%%%%%%%%%%%%%%%%%%%%%%%%%%%%%%%%%%%%%%

Similarly as in the case of Szilard's engine \cite{S29} we simplify the QMD scenario to a one-particle problem. Further,
we consider only two pairs of yes-no-properties of the particle:
\begin{itemize}
  \item Position: right or left (r/l),
  \item Speed: hot or cold (h/c).
\end{itemize}
This leads to a $4$-dimensional Hilbert space ${\mathcal H}={\mathbbm C}^4$ spanned by the four orthogonal
states $|rh\rangle, |rc\rangle,|lh\rangle,|lc\rangle$. For the L\"uders measurement we assume
\begin{equation}\label{M1}
  P_1=|rh\rangle\langle rh|,\quad P_2={\mathbbm 1}-P_1,\quad {\mathcal N}=\{1,2\}.
\end{equation}
As the conditional action we choose
\begin{equation}\label{M2}
  U_1=\left(
\begin{array}{cccc}
 0 & 0 & 1 & 0 \\
 0 & 1 & 0 & 0 \\
 1 & 0 & 0 & 0 \\
 0 & 0 & 0 & 1 \\
\end{array}
\right),\quad
U_2={\mathbbm 1}
\;.
\end{equation}
This means that, if the particle is found at the right hand side and being hot then
it is transferred to the left hand side without changing its speed:
\begin{equation}\label{M3}
  U_1 |rh\rangle =|lh\rangle
  \;.
\end{equation}
The action of $U_1$ onto the other three basis vectors is irrelevant since it models a conditional action
and will only be applied in the case where the first L\"uders measurement has the result ``yes" and yields the post measurement state $|rh\rangle$.
If the measurement result is ``no" then $U_2$ will be applied, i.~e., there will be no action.

Next we restrict the class ${\mathcal A}$ of admissible initial states to those of the form
\begin{equation}\label{M4}
  \rho =\frac{p}{2}\left( |rh\rangle\langle rh|+|lh\rangle\langle lh|\right)+
  \frac{1-p}{2}\left( |rc\rangle\langle rc|+|lc\rangle\langle lc|\right)
  \;,
\end{equation}
where $0<p<1$.
This means that initially the particle is in a mixed state with probability $p$ of being ``hot"
irrespective of its position. It follows that initially  the entropy will be
\begin{equation}\label{M5}
 S_0=S(\rho) =-\left( p\,\log \frac{p}{2}+(1-p)\log\frac{1-p}{2}\right)
 \;.
\end{equation}
The final state $\rho_1$ according to (\ref{QMD6}) will be
\begin{equation}\label{M6}
 \rho_1=p\,|rh\rangle\langle rh|+\frac{1-p}{2}\left( |rc\rangle\langle rc|+|lc\rangle\langle lc|\right)
\end{equation}
having the entropy
\begin{equation}\label{M7}
S_1=S(\rho_1) = -\left( p\,\log p+(1-p)\log\frac{1-p}{2}\right)
 \;.
\end{equation}
Comparison with (\ref{M5}) yields
\begin{equation}\label{M8}
S_1-S_0=-p\,\log 2  <0
\;,
\end{equation}
and hence the model is a proper QMD since the action of the demon leads to a decrease of the object system's entropy.

\begin{figure}[t]
\centering
\includegraphics[width=0.70\linewidth]{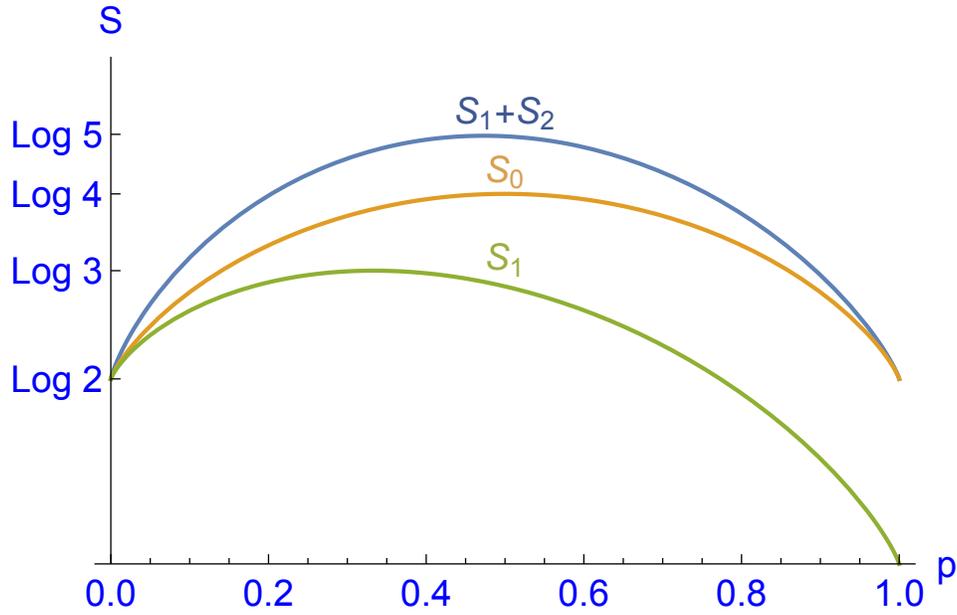}
\caption{Plot of various entropies of the simple QMD model as functions of the probability $p$:
The initial entropy $S_0$ of the total system ($=$ the initial entropy of the object system) (orange curve),
the final entropy $S_1+S_2$ of the total system (blue curve),
and the final entropy $S_1$ of the object system (green curve). We see that always $S_1<S_0$ but $S_1+S_2>S_0$
if $0<p<1$.}
\label{FIGENT}
\end{figure}

Our next aim is to construct a measurement dilation of the form (\ref{QMD11}) following the prescription
given in Appendix \ref{sec:E}.
Hence we choose ${\mathcal K}={\mathbbm C}^2$ with standard basis $\phi={1\choose 0}$ and $\psi={0\choose 1}$, and
\begin{equation}\label{M9}
  Q_1=P_\phi,\quad Q_2=P_\psi
\;.
\end{equation}
The linear operators in ${\mathcal H}\otimes {\mathcal K}={\mathbbm C}^4\otimes {\mathbbm C}^2 \cong {\mathbbm C}^4\oplus {\mathbbm C}^4$
will be represented by $2\times 2$-matrices the entries of which are $4\times 4$-matrices. This simplifies the calculation
of partial traces. With this convention we set
\begin{equation}\label{M10}
 V=\left(
\begin{array}{cccc|cccc}
 0 & 0 & 0 & 0 & 1 & 0 & 0 & 0 \\
 0 & 0 & 0 & 0 & 0 & 1 & 0 & 0 \\
 1 & 0 & 0 & 0 & 0 & 0 & 0 & 0 \\
 0 & 0 & 0 & 0 & 0 & 0 & 0 & 1 \\
 \hline
 0 & 0 & 0 & 0 & 0 & 0 & 1 & 0 \\
 0 & 1 & 0 & 0 & 0 & 0 & 0 & 0 \\
 0 & 0 & 1 & 0 & 0 & 0 & 0 & 0 \\
 0 & 0 & 0 & 1 & 0 & 0 & 0 & 0 \\
\end{array}
\right)
\,.
\end{equation}
One may confirm by direct calculation that with the above definitions
${\mathfrak D}_{{\mathcal K},\phi,V,Q}$ is indeed a measurement dilation of the considered Maxwell instrument.

Additionally, we will explicitly calculate the measurement dilation for admissible initial states
stepwise using the fact that all states will be diagonal in the standard basis of
$ {\mathbbm C}^4\oplus {\mathbbm C}^4$. First we note that
\begin{equation}\label{M11}
  \rho \otimes P_\phi= \mbox{diag}\left( \frac{p}{2},\frac{1-p}{2}, \frac{p}{2},\frac{1-p}{2},0,0,0,0\right)
  \;.
\end{equation}
Since $V\left( \rho \otimes P_\phi\right) V^\ast$ is already diagonal we obtain
\begin{eqnarray}\label{M12a}
\rho_{12}&=& V\left( \rho \otimes P_\phi\right)V^\ast\\
\label{M12b}
&=&
\sum_{n=1}^2 \left({\mathbbm 1}\otimes Q_n\right)\, V\left( \rho \otimes P_\phi\right)V^\ast\,\left({\mathbbm 1}\otimes Q_n\right)\\
 \label{M12c}
 &=&
 \mbox{diag}\left(0,0, \frac{p}{2},0,0,\frac{1-p}{2}, \frac{p}{2},\frac{1-p}{2}\right)
  \;.
\end{eqnarray}
From this we obtain the partial trace $\rho_1$ as the sum of the two diagonal blocks of $\rho_{12}$:
\begin{equation}\label{M13}
  \rho_1= \mbox{Tr}_{\mathcal K}\,\rho_{12}=\mbox{diag}\left( 0,\frac{1-p}{2},p,\frac{1-p}{2}\right)
\end{equation}
in accordance with (\ref{M6}) and its entropy (\ref{M7}). Analogously, the final state of the demon is obtained
by taking the traces of the block matrices and has the form
\begin{equation}\label{M15}
  \rho_2= \mbox{Tr}_{\mathcal H}\,\rho_{12}=\mbox{diag}\left( \frac{p}{2},1-\frac{p}{2}\right)
\end{equation}
with entropy
\begin{equation}\label{M16}
  S_2=S(\rho_2)=-\left( \frac{p}{2}\,\log \frac{p}{2}+(1-\frac{p}{2})\,\log \left(1-\frac{p}{2}\right)\right)
  \;.
\end{equation}
This leads to
\begin{equation}\label{M17}
  S_1<S_0,\quad \mbox{but } S_1+S_2>S_0
  \;,
\end{equation}
see Figure \ref{FIGENT}, and hence the decrease of entropy of the object system is overcompensated by the increase
of the demon's entropy in our example.

A remarkable detail of our example is the fact that the state of the combined system after the interaction
\begin{equation}\label{M18}
 V\left( \rho \otimes P_\phi\right)V^\ast
\end{equation}
commutes with all projections ${\mathbbm 1}\otimes Q_n$ and hence the entropy increase due to the L\"uders measurement
vanishes. The final entropy increase is completely due to the separation of the total state into reduced
states of the subsystems.
It has been argued against Szilard's principle that there are also reversible measurements
and hence this principle alone is not sufficient to defense the $2^{nd}$ law against the Maxwell's demon objection,
see \cite{B82}, chapter $5$.
Our example yields a counter argument closely related to Zurek's consideration of mutual information \cite{Z84}:
In the quantum case there are also entropy costs of state separation
that might suffice to compensate the entropy decrease of the object system even if the measurement is reversible (adiabatic).

%%%%%%%%%%%%%%%%%%%%%%%%%%%%%%%%%%%%%%%%%%
\section{Classical conditional action}
\label{sec:C}
%%%%%%%%%%%%%%%%%%%%%%%%%%%%%%%%%%%%%%%%%%
It will be instructive to investigate the classical counterpart of the conditional action relative to a (L\"uders) measurement
introduced in Section \ref{sec:QMD}. To this end we consider probability distributions
\begin{equation}\label{C1}
  p:{\mathcal I}\rightarrow [0,1]
\end{equation}
defined on a finite set ${\mathcal I}$ of elementary events and subject to the condition
\begin{equation}\label{C2}
  \sum_{i\in{\mathcal I}}p_i=1
  \;.
\end{equation}
A ``measurement" will be represented by a partition of ${\mathcal I}$, i.~e., a disjoint union
\begin{equation}\label{C3}
  {\mathcal I} = \biguplus_{j\in{\mathcal J}}I_j
  \;.
\end{equation}

As usual, we define the Shannon entropy \cite{S48}, up to a factor $\log 2$, by
\begin{equation}\label{C4}
 H(p)\equiv -\sum_{i\in{\mathcal I}}p_i\,\log p_i
 \;.
\end{equation}
Then a ``classical conditional action" relative to the measurement $\left(I_j\right)_{j\in{\mathcal J}}$
will be defined by a map
\begin{equation}\label{C5}
 \phi:{\mathcal I} \rightarrow {\mathcal I}
 \;,
\end{equation}
that is injective on the subsets $I_j$, i.~e., if $i_1, i_2\in I_j$ for some ${j\in{\mathcal J}}$
and $i_1\neq i_2$ then $\phi(i_1)\neq \phi(i_2)$ holds.
Each conditional action gives rise to a new probability distribution $q:{\mathcal I}\rightarrow [0,1]$ defined by
\begin{equation}\label{C6}
 q_i\equiv \sum_{\phi(k)=i}p_k
 \;,
\end{equation}
that has, in contrast to the quantum case, {\em always} a lower (or the same) entropy:
\begin{equation}\label{C7}
  H(q)\le H(p)
  \;.
\end{equation}
\noindent{\bf Proof} of Eq.~\ref{C7}:
If $\phi$ is a global bijection then (\ref{C7}) is satisfied with equality.
Now assume that exactly two events are mapped onto the same one, say,  $\phi(1)=\phi(2)=i$
and $p_1, p_2 >0$.  Then we conclude, for $j=1,2,$
\begin{eqnarray}
\label{C8a}
  \log p_j &<& \log (p_1+p_2), \\
  \label{C8b}
  - \log p_j &>& -\log (p_1+p_2), \\
  \label{C8c}
 -p_1\log p_1-p_2\log p_2& >& -(p_1+p_2)\log(p_1+p_2)\\
 \label{C8d}
 &=&-q_i\,\log q_i
 \;,
\end{eqnarray}
which means that the fusion of two probabilities $p_1$ and $p_2$ to $q_i$ decreases the corresponding
term of the entropy. From this the general case follows by induction. \hfill$\Box$\\

We will give an elementary example. Let ${\mathcal I}=\{1,2,3,4,5,6\}$ denote the numbers of a die
and $p_i=1/6$ their probabilities. The measurement detects whether the dice roll is low or high,
corresponding to the partition ${\mathcal I}=I_1 \uplus I_2 = \{1,2,3\} \uplus \{4,5,6\}$.
If the dice roll is low, the die is flipped so that the new roll is high. If the dice roll is already high, nothing is done.
This describes the conditional action
\begin{equation}\label{C9}
  \phi(i)=\left\{
 \begin{array}{r@{\quad \mbox{if} \quad}l}
i & i\in I_2,\\
 7-i & i\in I_1.
 \end{array}
 \right.
\end{equation}
The new probability distribution $q$ generated by the conditional action will be given by
$q_1=q_2=q_3=0$ and $q_4=q_5=q_6=1/3$. It has the entropy $H(q)=\log 3 <H(p)=\log 6$,
in accordance with (\ref{C7}).

Returning to the general case we will define the analogue of the ``measurement dilation" considered in Section \ref{sec:QMD}.
The first step is to consider the extended event space
\begin{equation}\label{C10}
  \Omega= {\mathcal I}\times  {\mathcal J}
\end{equation}
and a fixed initial value $j_0\in {\mathcal J}$. This means that the initial distribution $P_2:{\mathcal J}\rightarrow [0,1]$
is concentrated on the value $j_0$ and hence has vanishing entropy, $H(P_2)=0$.

Define the injective map $\Phi:{\mathcal I}\times\{j_0\}\rightarrow  \Omega$ defined by
\begin{equation}\label{C11}
 \Phi(i,j_0)\equiv (\phi(i),{\sf j}(i))
 \;,
\end{equation}
where we have written $j={\sf j}(i)$ if $i\in I_j$. The injectivity of
$\Phi$ follows since $i_1\neq i_2$ and $i_1,i_2\in I_j$ for some $j\in {\mathcal J}$ implies
$\phi(i_1,j_0)\neq \phi(i_2,j_0)$ by the assumption that $\phi$ is injective on $I_j$. If  $i_1,i_2$ lie in different
sets $I_j$ then ${\sf j}(i_1) \neq {\sf j}(i_2))$.
Hence $\Phi$ can be extended to a bijective map $\bar{\Phi}:\Omega\rightarrow\Omega$
that is the analogue of the unitary operator $V$ introduced in Eq.~(\ref{QMD9}).

$\Phi$ maps $p$ onto a new probability distribution $Q$
on $\Omega$ defined by
\begin{equation}\label{C12}
  Q(k,j)=\left\{
 \begin{array}{r@{\quad : \quad}l}
p_i & \mbox{if }\Phi(i,j_0)=(k,j),\\
 0 & \mbox{else},
 \end{array}
 \right.
\end{equation}
with the same entropy, $H(Q)=H(p)$. Let $Q_1$ denote the first marginal distribution of $Q$  given by
\begin{equation}\label{C13}
  Q_1(k)=\sum_{j\in{\mathcal J}}Q(k,j)
  \;,
\end{equation}
and, analogously,
\begin{equation}\label{C14}
  Q_2(j)=\sum_{k\in{\mathcal I}}Q(k,j)
  \;.
\end{equation}
Then it can be shown that $Q_1$ coincides with the distribution $q$ defined above. The proof uses
\begin{eqnarray}\label{C15a}
 Q_1(k)&\stackrel{(\ref{C13})}{=}&\sum_{j\in{\mathcal J}}Q(k,j)\\
 &\stackrel{(\ref{C12})}{=}& \sum_{j\in{\mathcal J},\Phi(i,j_0)=(k,j) } p_i\\
 &\stackrel{(\ref{C11})}{=}& \sum_{j\in{\mathcal J},(\phi(i),{\sf j}(i))=(k,j) } p_i\\
 &=& \sum_{\phi(i)=k}p_i\\
 &\stackrel{(\ref{C6})}{=}& q(k)
 \;.
\end{eqnarray}
By the subadditivity of the Shannon entropy, see \cite{NC00} Theorem 11.3 (4), we have $H(Q_1)+H(Q_2)\ge H(Q)$
and hence $H(Q_2)\ge H(Q)-H(Q_1)=H(p)-H(q)$. This means that the entropy decrease $H(q)-H(p)<0$
due to the conditional action is (over)compensated by entropy increase of $H(Q_2)-H(P_2)=H(Q_2)$, analogously to the quantum case.

In order to illustrate the measurement dilation for the above example, we first note that ${\mathcal J}=\{1,2\}$
can be viewed as a kind of memory of whether the die has been flipped ($j=2$) or not ($j=1$).
Let $j_0\equiv1$, then the map $\Phi$ is given by
\begin{eqnarray}\nonumber
&&\Phi(1,1)=(6,2),\;\Phi(2,1)=(5,2),\;\Phi(3,1)=(4,2),\\
\nonumber
&&\Phi(4,1)=(4,1),\;\Phi(5,1)=(5,1),\;\Phi(6,1)=(6,1).\\
&&\label{C16}
\end{eqnarray}
The  resulting probability distribution $Q$ satisfies
\begin{eqnarray}\nonumber
Q(4,1)&=&Q(5,1)=Q(6,1)\\
\label{C17}
&=&Q(4,2)=Q(5,2)=Q(6,2)=1/6
\;,
\end{eqnarray}
and vanishes for other events. Hence $H(Q)=\log 6$.
The marginal distributions are obtained as $Q_1=q$ and $Q_2(1)=Q_2(2)=1/2$.
Hence $H(Q_1)=\log 3$ and $H(Q_2)=\log 2$. The latter exactly compensates the entropy decrease
$H(q)-H(p)=-\log 2$ due to the conditional action.

%%%%%%%%%%%%%%%%%%%%%%%%%%%%%%%%%%%%%%%%%%
\section{Summary}
\label{sec:SU}
%%%%%%%%%%%%%%%%%%%%%%%%%%%%%%%%%%%%%%%%%%

We have given an explanation of the apparently paradoxical entropy decrease of a quantum system caused by the external
intervention analogous to but more general than Maxwell's demon. This explanation
follows Szilard's principle \cite{S29} and its quantum version given by Zurek \cite{Z84} in so far as it includes
the demon's state into the entropy balance. But we extend these approaches by introducing the concept of  ``conditional action"
and its mathematical description in terms of a ``Maxwell instrument". The quantum-mechanical description of the demon
can then be accomplished by using tools from quantum measurement theory \cite{BLPY16}, especially the ``measurement dilation"
of a Maxwell instrument. The entropy decrease due to the conditional action of Maxwell's demon thus appears as a special
case of the entropy decrease due to a non-L\"uders measurement and has an analogous explanation,
see \cite{L73}, \cite{BLM96} or \cite{SG20}. Of course, we have not shown that {\em all} physical realizations
of Maxwell's demon would be compatible with a tentative $2^{nd}$ law, but only those described by measurement dilations.

The relation of our explanation to the Landauer/Bennett principle proves to be ambivalent.
On the one hand there is no contradiction: If the conditional action is intended to be part of a cyclic
process it would be necessary to reset the state of the demon to its initial value. This is only possible
by another conditional action performed by a second demon and ends up with an increased entropy of the second
demon's state. But on the other hand it would not be entirely appropriate to call this process an ``erasure of memory"
since in our approach the function of the demon cannot be reduced to a mere memory, but also includes
the role of a measuring device and of a control unit for the conditional action. Moreover, the reset of the demon's state
was motivated by getting started a cyclic process. If this reset necessarily increases the entropy of some other part of the
environment, this simply means that it has not achieved its goal and hence is superfluous. From this perspective
the  Landauer/Bennett principle appears as a possible supplement to Szilard's principle
but can hardly be viewed as ``the ultimate reason for the entropy increase" \cite{Z84}.

It has been argued \cite{EN99} that current explanations of Maxwell's demon using principles connecting information and entropy
are not yet based on firm grounds. It is therefore worth mentioning that our approach does not rely on concepts from information theory,
notwithstanding the frequent citation of a textbook \cite{NC00} on quantum information theory and the use of von Neumann entropy.
One may object, what is information anyway, if not the result of measurements used to trigger conditional action?
But what one is actually concerned with here is the methodological distinction between specialization and generalization.
It may be possible to introduce new concepts that fit specific situations without extending the theory in question.
However, this must be strictly separated from the situation where new terms and laws are required to generalize the theory.
Conceptual parsimony can be helpful to clearly distinguish between these two cases.

\appendix

%%%%%%%%%%%%%%%%%%%%%%%%%%%%%%%%%%%%%%%%%%
\section{Characterization of Maxwell instruments}
\label{sec:M}
%%%%%%%%%%%%%%%%%%%%%%%%%%%%%%%%%%%%%%%%%%

There exists a so-called statistical duality between states and observables, see \cite{BLPY16}, chapter 23.1.
In the finite-dimensional case $B({\mathcal H})$ can be identified with its dual space $B({\mathcal H})^\ast$
by means of the Euclidean scalar product $\mbox{Tr}\,(A\,B)$. Physically, we may distinguish between the two
spaces in the sense that $B({\mathcal H})$ is spanned by the subset of statistical operators representing {\em states}
and $B({\mathcal H})^\ast$  is spanned by the subset of operators with eigenvalues in the interval $[0,1]$ representing
{\em effects}. Effects describe yes-no-measurements including the subset of projectors, which are the extremal points of the
convex set of effects, see \cite{BLPY16}.

Every operation $A:B({\mathcal H})\rightarrow B({\mathcal H})$, viewed as a transformation of states (Schr\"odinger picture)
gives rise to the dual operation $A^\ast:B({\mathcal H})^\ast\longrightarrow B({\mathcal H})^\ast$
viewed as a transformation of effects (Heisenberg picture).
Reconsider the representation  (\ref{OI2}) of the operation $A$ by means of the Kraus operators $A_i$. Then the dual
operation $A^\ast$ has the corresponding representation
\begin{equation}\label{M1}
 A^\ast(X)= \sum_{i\in{\mathcal I}}A_i^\ast\,X\,A_i
  \;,
\end{equation}
for all $X\in B({\mathcal H})^\ast$.

Similarly, every instrument ${\mathfrak I}$ gives rise to a dual instrument
${\mathfrak I}^\ast:{\mathcal N}\times B({\mathcal H})^\ast\longrightarrow B({\mathcal H})^\ast$
defined by
\begin{equation}\label{M2}
 {\mathfrak I}^\ast(n)(X)\equiv  {\mathfrak I}(n)^\ast(X)
\end{equation}
for all $n\in {\mathcal N}$ and $X\in B({\mathcal H})^\ast$. The condition that the total operation (\ref{OI4}) will
be trace-preserving translates into
\begin{equation}\label{M3}
 {\mathfrak I}^\ast({\mathcal N})({\mathbbm 1})=\sum_{n\in{\mathcal N}} {\mathfrak I}^\ast(n)({\mathbbm 1})={\mathbbm 1}
 \;.
\end{equation}
Thus every dual instrument yields a resolution of the identity by means of effects $F_n= {\mathfrak I}^\ast(n)({\mathbbm 1})$,
and hence to a generalized observable in the sense of a {\em positive operator-valued measure} (POM) \cite{BLPY16}.
The traditional notion of ``sharp"  observables represented by self-adjoint operators corresponds to the special case
of a projection-valued measure $P_n,\;n\in {\mathcal N},$ satisfying $\sum_{n\in{\mathcal N}}P_n={\mathbbm 1}$.

An operation $A:B({\mathcal H})\rightarrow B({\mathcal H})$ will be called {\em pure} iff it maps rank one operators
onto rank one operators. Physically, this means that a pure operation maps pure states onto pure states, up to a positive factor.
If the representation  (\ref{OI2}) of $A$  can be reduced to a single Kraus operator, i.~e.,
\begin{equation}\label{M4}
  A(\rho) = A_1\,\rho\,A_1^\ast
  \;,
\end{equation}
then $A$ will be a pure operation. Conversely, every pure operation has a representation of the form (\ref{M4}), as can be shown by means of
lemma 7.8 in \cite{E72} (note that this reference does not require complete positivity for operations).
{\em Pure instruments} are defined analogously. Note that Maxwell instruments will be pure since they consist of pure operations
according to (\ref{QMD5}).

Then we can formulate the following characterization of Maxwell instruments:
%%%%%%%%%%%%%%%%%%%%%%%%%%%%%%%%%%%%%%%%%%%%%%%%%%%%%%%%%%%%%%%%%%%%%%%%%%%%%%%%%%%%%%%%%%%%%%%%%%%%%%%%%%%%%%%%%%%%%%%%%%%%%%%%%%%
\begin{prop}\label{PC}
 An instrument  ${\mathfrak I}$ will be a Maxwell instrument iff it is pure and gives rise to a sharp observable,
 i.~e., $P_n\equiv {\mathfrak I}^\ast(n)({\mathbbm 1})$
 will be a projection for all $n\in {\mathcal N}$.
\end{prop}
%%%%%%%%%%%%%%%%%%%%%%%%%%%%%%%%%%%%%%%%%%%%%%%%%%%%%%%%%%%%%%%%%%%%%%%%%%%%%%%%%%%%%%%%%%%%%%%%%%%%%%%%%%%%%%%%%%%%%%%%%%%%%%%%%%%%%
{\bf Proof} \\
``only-if-part": As remarked above, a Maxwell instrument ${\mathfrak I}$ will be pure and, according to (\ref{QMD5}), its dual has the representation
\begin{equation}\label{M6}
  {\mathfrak I}^\ast (n)(X)=P_n\,U^\ast\,X\, U\,P_n
  \;,
\end{equation}
for all $n\in {\mathcal N}$ and $X\in B({\mathcal H})^\ast$. Hence
\begin{equation}\label{M7}
  {\mathfrak I}^\ast (n)({\mathbbm 1})=P_n\,U^\ast\,{\mathbbm 1}\, U\,P_n= P_n
  \;,
\end{equation}
will be a projector for all $n\in {\mathcal N}$.\\
``if-part":  If ${\mathfrak I}(n)$ is pure it has the representation (\ref{M4}) and hence
\begin{equation}\label{M8}
 {\mathfrak I}^\ast (n)(X)= A_n^\ast\,X\,A_n
 \;,
\end{equation}
for all $n\in {\mathcal N}$ and $X\in B({\mathcal H})^\ast$. Let $A_n^\ast=Q_n\,U_n^\ast$ be the polar decomposition
of $A_n^\ast$ such that $Q_n$ is positively semi-definite and $U_n^\ast$ unitary. It follows by assumption that
\begin{equation}\label{M9}
 {\mathfrak I}^\ast (n)({\mathbbm 1})= A_n^\ast\,A_n= \left( Q_n\,U_n^\ast\right)\left(U_n\, Q_n\right)= (Q_n)^2
\end{equation}
is a projector for all $n\in {\mathcal N}$ and hence we may set $Q_n=P_n$ and obtain
\begin{equation}\label{M10}
   {\mathfrak I}(n)(\rho) = A_n\,\rho\,A_n^\ast = U_n\,P_n\,\rho\,P_n\,U_n^\ast
   \;,
\end{equation}
for all $n\in {\mathcal N}$ and $\rho\in B({\mathcal H})$. This shows that ${\mathfrak I}$ is a Maxwell instrument
and completes the proof of Proposition \ref{PC}.
 \hfill$\Box$\\

%%%%%%%%%%%%%%%%%%%%%%%%%%%%%%%%%%%%%%%%%%
\section{Conditional action decreases entropy}
\label{sec:P}
%%%%%%%%%%%%%%%%%%%%%%%%%%%%%%%%%%%%%%%%%%
\noindent{\bf Proof} of Proposition \ref{propMAX}: Define
\begin{equation}\label{P1}
 p_n \equiv \mbox{Tr}\left( \rho\,P_n\right)
 \;,
\end{equation}

\begin{equation}\label{P3}
 \rho_n \equiv \frac{1}{p_n}\,P_n\,\rho\,P_n
 \;,
\end{equation}
and
\begin{equation}\label{P4}
 \widetilde{\rho}_n \equiv \frac{1}{p_n}\,U_n\,P_n\,\rho\,P_n\,U_n^\ast = U_n\,\rho_n\,U_n^\ast
 \;,
\end{equation}
for all $n\in{\mathcal N}$. Obviously,
\begin{equation}\label{P5}
  S(\widetilde{\rho}_n)=S(\rho_n)
  \;.
\end{equation}
Since the $\rho_n$ have orthogonal support, theorem 11.8 (4) of \cite{NC00} can be applied and yields:
\begin{equation}\label{P6}
\widetilde{S}_1= S\left(\sum_{n\in{\mathcal N}}p_n\rho_n\right) =\sum_{n\in{\mathcal N}} p_n S(\rho_n)+ H(p)
 \;,
\end{equation}
where $H(p)$ is the Shannon entropy, see (\ref{C4}).
Analogously, theorem 11.10 of \cite{NC00} yields
\begin{eqnarray}\label{P8}
 S_1&=& S\left(\sum_{n\in{\mathcal N}}p_n\widetilde{\rho}_n\right) \le\sum_{n\in{\mathcal N}} p_n S(\widetilde{\rho}_n)+ H(p)\\
  &\stackrel{(\ref{P5})}{=}&\sum_{n\in{\mathcal N}} p_n S({\rho}_n)+ H(p)\\
   &\stackrel{(\ref{P6})}{=}&S\left(\sum_{n\in{\mathcal N}}p_n\rho_n\right)=\widetilde{S}_1
  \;,
\end{eqnarray}
which completes the proof of  Proposition \ref{propMAX}. \hfill$\Box$\\

If the initial state $\rho$ and the family of projections $P_n,\,n\in{\mathcal N},$ is given, one may ask
which choice of the unitary operators $U_n,\,n\in{\mathcal N},$ would minimize the entropy
$S_1=S\left(\sum_{n\in{\mathcal N}} U_n\,P_n\,\rho P_n\,U_n^\ast\right)$? We conjecture
the following result.

Let $\left(\psi_\mu\right)_{\mu\in{\mathcal M}}$ be an orthonormal basis in ${\mathcal H}$ and
$\left(\phi_\mu^{(n)}\right)_{\mu=1,\ldots,d_n}$ an eigenbasis of $P_n\,\rho\,P_n$ such that
\begin{equation}\label{P9}
 P_n\,\rho\,P_n\,\phi_\mu^{(n)}=r_\mu^{(n)}\,\phi_\mu^{(n)}
\end{equation}
for all $\mu=1,\ldots,d_n$ and $n\in{\mathcal N}$.
We assume that the order of the indices $\mu$ is chosen such that the eigenvalues of $P_n\,\rho\,P_n$
are monotonically decreasing:
\begin{equation}\label{P10}
 r_1^{(n)}\ge r_2^{(n)}\ge\ldots \ge r_{d_n}^{(n)}
\end{equation}
for all $n\in{\mathcal N}$. Then an optimal choice of the  $U_n$ is given by the conditions
\begin{equation}\label{P11}
  U_n\,\phi_\mu^{(n)}=\psi_\mu
\end{equation}
for all $\mu=1,\ldots,d_n$ and $n\in{\mathcal N}$. This means that the $U_n$ merge the
eigenspaces of $P_n\,\rho\,P_n$  as much as possible such that the largest corresponding eigenvalues
are added thereby decreasing the entropy of the state. The above choice is not unique since, e.~g.,
global permutations of the eigenvalues do not change the entropy.

Of course, it is not clear in general whether this decrease of entropy leads to $S_1<S_0$. Only
in the latter case we would call the resulting Maxwell instrument ``demonic". If the choice of the
$P_n,\,n\in{\mathcal N},$ also remains open the problem becomes trivial:
Upon choosing the $P_n,\,n\in{\mathcal N},$ one-dimensional the above optimal choice
of the $U_n$ yields a pure state with vanishing entropy,
as in the case of erasure of $N$ qubits in Section \ref{sec:ER}.

%%%%%%%%%%%%%%%%%%%%%%%%%%%%%%%%%%%%%%%%%%
\section{Explicit construction of a measurement dilation for a Maxwell instrument}
\label{sec:E}
%%%%%%%%%%%%%%%%%%%%%%%%%%%%%%%%%%%%%%%%%%

Let a Maxwell instrument of the form (\ref{QMD5}) be given, i.~e.,
\begin{equation}\label{E1}
 \mathfrak{J}(n)(\rho)= U_n\,P_n\,\rho P_n\,U_n^\ast,\quad n\in{\mathcal N}
 \;.
\end{equation}
Following \cite{NC00} we want to explicitly construct a measurement dilation of $\mathfrak{J}$ of the form (\ref{QMD11}).
To this end we choose ${\mathcal K}={\mathbbm C}^{\mathcal N}$ and an orthonormal basis $|n\rangle_{n\in{\mathcal N}}$
in ${\mathcal K}$.
Let $\phi\in{\mathcal K}$ be one of these basis vectors, say,
$\phi=|1\rangle$.

Further, let $\check{P}_n$ denote the eigenspace of the projector $P_n$
corres\-ponding to the eigenvalues $1$
and $|ni\rangle_{i=1,\ldots, \dim \check{P}_n}$ some orthonormal bases in $\check{P}_n$
such that
\begin{equation}\label{E1a}
 P_n = \sum_i |ni\rangle \langle ni|\;,\quad
 \mbox{for all }  n\in{\mathcal N}
 \;.
\end{equation}
Moreover, let $\check{Q}_n\equiv {\mathcal H}\otimes |n\rangle$ and $Q_n=|n\rangle\langle n|$ denote the projector onto
the one-dimensional subspace spanned by $|n\rangle$ for all $n\in{\mathcal N}$.
We define a linear map $V_1:\check{Q}_1 \rightarrow {\mathcal H}\otimes{\mathcal K}$ by
\begin{equation}\label{E2}
  V_1 |ni1\rangle \equiv  V_1\left( |ni\rangle \otimes |1\rangle\right)
  \equiv U_n  |ni\rangle  \otimes |n\rangle = \left( U_n \otimes {\mathbbm 1}\right) |nin\rangle
  \end{equation}
where $i=1,\ldots, \dim \check{P}_n$ and $n\in{\mathcal N}$.

%%%%%%%%%%%%%%%%%%%%%%%%%%%%%%%%%%%%%%%%%%%%%%%%%%%%%%%%%%%%%%%%%%%%%%%%%%%%%%%%%%%%%%%%%%%%%%%%%%%%%%%%%%%%%%%
\begin{lemma}\label{L1}
 The map $V_1:\check{Q}_1 \rightarrow {\mathcal H}\otimes{\mathcal K}$ is a partial iso\-metry, i.~e.,
 satisfies $V_1^\ast\,V_1={\mathbbm 1}_{\check{Q}_1}$.
\end{lemma}
%%%%%%%%%%%%%%%%%%%%%%%%%%%%%%%%%%%%%%%%%%%%%%%%%%%%%%%%%%%%%%%%%%%%%%%%%%%%%%%%%%%%%%%%%%%%%%%%%%%%%%%%%%%%%%%%%%%%%

Proof: Let $|mj1\rangle$ and $|ni1\rangle$ be two arbitrary vectors of the orthonormal basis of $\check{Q}_1$
obtained from the orthonormal basis of ${\mathcal H}$ considered above.
Then we conclude
\begin{eqnarray}
\nonumber
  \left\langle   mj1 \left| V_1^\ast\,V_1\right | ni1\right\rangle
  &\stackrel{(\ref{E2})}{=}& \left\langle   mjm \left| \left(U_m^\ast \otimes {\mathbbm 1} \right)\left(U_n \otimes {\mathbbm 1} \right)\right| nin\right\rangle \\
  \label{E3a} &&\\
   &=& \left\langle mj\left| U_m^\ast\,U_n\right| ni\right\rangle \underbrace{\langle m|n\rangle}_{\delta_{mn}}\\
   \label{E3b}
   &=& \left\langle n j\left| U_n^\ast\,U_n\right| ni\right\rangle\,\delta_{mn}\\
   \label{E3c}
    &=& \left\langle n j\left| {\mathbbm 1}\right| ni\right\rangle\,\delta_{mn}\\
    \label{E3d}
   &=& \delta_{ij}\,\delta_{mn}\\
   \label{E3e}
   &=&  \left\langle   mj1 \left|{\mathbbm 1}_{\check{Q}_1}\right | ni1\right\rangle
  \;,
\end{eqnarray}
which completes the proof of Lemma \ref{L1}. \hfill$\Box$\\

Next we extend the partial isometry $V_1$ to a unitary operator
$V:{\mathcal H}\otimes{\mathcal K}\rightarrow {\mathcal H}\otimes{\mathcal K}$.
This completes the definition of the quantities ${\mathcal K},\phi,V,Q$ required for the measurement dilation.
It remains to show that $\mathfrak{J}={\mathfrak D}_{{\mathcal K},\phi,V,Q}$. To this end we write
\begin{equation}\label{E4}
 \rho=\sum_{\ell,i,m,j}|\ell i\rangle \langle \ell i| \rho | mj\rangle \langle mj|
\end{equation}
and
\begin{equation}\label{E5}
 \rho\otimes P_\phi=\sum_{\ell,i,m,j}|\ell i 1\rangle \langle \ell i| \rho | mj\rangle \langle mj1|
 \;.
\end{equation}
Further,
\begin{eqnarray}
\nonumber
&&
 V\left(\rho\otimes P_\phi\right)V^\ast\\
 \label{E6a}
 &\stackrel{(\ref{E5})}{=}& \sum_{\ell,i,m,j}V\,|\ell i 1\rangle \langle \ell i| \rho | mj\rangle \langle mj1|\,V^\ast \\
  \nonumber
   &\stackrel{(\ref{E2})}{=}&\sum_{\ell,i,m,j}\left(U_\ell\otimes {\mathbbm 1} \right)|\ell i \ell\rangle
   \langle \ell i| \rho | mj\rangle
   \langle mjm|\left(U_m^\ast\otimes {\mathbbm 1} \right)\\
   \label{E6b}
   &&\\
   \label{E6c}
   &=& \sum_{\ell,i,m,j}
   \left(
   U_\ell |\ell i\rangle
    \langle \ell i| \rho | mj\rangle
   \langle mj|U_m^\ast
   \right)
   \otimes |\ell\rangle\langle m|
   \;.
\end{eqnarray}
Using
\begin{equation}\label{E7}
Q_n|\ell\rangle\langle m|Q_n=\delta_{\ell n}\,\delta_{mn}\,Q_n
\;,
\end{equation}
(\ref{E6c}) implies
\begin{eqnarray}
\nonumber
  &&\left({\mathbbm 1}\otimes Q_n \right)  V\left(\rho\otimes P_\phi\right)V^\ast \left({\mathbbm 1}\otimes Q_n \right)\\
  \label{E8}
  &=& \sum_{i,j}\left(
  U_n|ni\rangle   \langle n i| \rho |nj\rangle\langle nj| U_n^\ast
  \right)
  \otimes Q_n
  \;,
\end{eqnarray}
and
\begin{eqnarray}
\nonumber
&& {\mathcal D}_{{\mathcal K},\phi,V,Q}(\rho)\\
\nonumber
  &=&\mbox{Tr}_{\mathcal K}
  \left(\left({\mathbbm 1}\otimes Q_n \right)  V\left(\rho\otimes P_\phi\right)V^\ast \left({\mathbbm 1}\otimes Q_n \right)\right)\\
  \label{E9a}
  &\stackrel{(\ref{E8})}{=}& \sum_{i,j}
  U_n|ni\rangle   \langle n i| \rho |nj\rangle \langle nj|U_n^\ast
  \;,
\end{eqnarray}
since $\mbox{Tr}\, Q_n=1$ for all $n\in{\mathcal N}$.
The latter expression equals
\begin{eqnarray}
\nonumber
  \mathfrak{J}(n)(\rho)&=& U_n\,P_n\,\rho\,P_n\,U_n^\ast \\
   &\stackrel{(\ref{E1a})}{=}& \sum_{i,j} U_n |ni\rangle \langle n i| \rho |nj\rangle\langle nj| U_n^\ast
   \;,
\end{eqnarray}
thereby proving that the above construction is a correct measurement dilation of ${\mathfrak J}$.\\

Next we calculate the reduction of the final state to the demon subsystem and obtain
\begin{eqnarray}
\nonumber
  \rho_2&\equiv&\mbox{Tr}_{\mathcal H}
  \sum_{n\in{\mathcal N}}
  \left({\mathbbm 1}\otimes Q_n \right)  V\left(\rho\otimes P_\phi\right)V^\ast \left({\mathbbm 1}\otimes Q_n \right)\\
  \label{E10a}
  &\stackrel{(\ref{E8})}{=}& \sum_{n\in{\mathcal N}}
 \mbox{Tr}\left(P_n\,\rho\,P_n \right)\,Q_n \stackrel{(\ref{P1})}{=} \sum_{n\in{\mathcal N}} p_n\,Q_n
  \;.
\end{eqnarray}
The corresponding entropy amounts to the Shannon entropy
\begin{equation}\label{E11}
S_2=S(\rho_2)=-\sum_{n\in{\mathcal N}}p_n\,\log p_n \stackrel{(\ref{C4})}{=} H(p)\ge 0
\;.
\end{equation}

In connection with the Szilard principle the following result is interesting:
%%%%%%%%%%%%%%%%%%%%%%%%%%%%%%%%%%%%%%%%%%%%%%%%%%%%%%%%%%%%%%%%%%%%%%%%%%%%%%%%%%%%%%%%%%%%%%%%%%%%%%%%%%%%%%%%%%%%%%%%%%%%%%%%%%%%%%%%%%%%%%%%%%%%%%%%%%%%%%%%%%
\begin{prop}\label{propML}
The total entropy of the composed state after the measurement dilation ${\mathfrak D}_{{\mathcal K},\phi,V,Q}$
constructed above exceeds (or equals) the entropy of the state after the corresponding L\"uders operation,
\begin{equation}\label{ML1}
\widetilde{S}_1 \le S_1+S_2
\;.
\end{equation}
\end{prop}
%%%%%%%%%%%%%%%%%%%%%%%%%%%%%%%%%%%%%%%%%%%%%%%%%%%%%%%%%%%%%%%%%%%%%%%%%%%%%%%%%%%%%%%%%%%%%%%%%%%%%%%%%%%%%%%%%%%%%%%%%%%%%%%%%%%%%%%%%%%%%%%%%%%%%%%%%%%%%%%%%%
\noindent{\bf Proof} of Proposition \ref{propML}:
With the definitions (\ref{P1}) -- (\ref{P4}) we conclude from the concavity of the von Neumann entropy, see (11.86) in \cite{NC00},
\begin{equation}\label{E12}
 \sum_{n\in{\mathcal N}}p_n\,S(\rho_n)\stackrel{(\ref{P5})}{=}\sum_{n\in{\mathcal N}}p_n\,S(\widetilde{\rho}_n)
 \le S\left(\sum_{n\in{\mathcal N}}p_n\,\widetilde{\rho}_n
 \right)=S_1
 \;.
\end{equation}
This further implies
\begin{equation}\label{E13}
 \widetilde{S}_1- S_2\stackrel{(\ref{E11})}{=} \widetilde{S}_1-H(p)\stackrel{(\ref{P6})}{=} \sum_{n\in{\mathcal N}}p_n\,S(\rho_n)
 \stackrel{(\ref{E12})}{\le} S_1
 \;,
\end{equation}
and (\ref{ML1}) immediately follows. \hfill$\Box$\\

%%%%%%%%%%%%%%%%%%%%%%%%%%%%%%%%%%%%%%%%%%
\section{The Szilard engine revisited}
\label{sec:S}
%%%%%%%%%%%%%%%%%%%%%%%%%%%%%%%%%%%%%%%%%%
We will reconsider the Szilard engine, but in contrast to the simplified model in section \ref{sec:EM},
adopt a more realistic description of the one-molecule gas and the isothermal expansion after position measurement.
In doing so, we will stick to \cite{Z84} as far as possible, but emphasize the differences to the present approach.

In classical thermodynamics there are various equivalent formulations of the $2^{nd}$ law including
the impossibility of a perpetuum mobile of the second kind. This is a cyclic process transforming heat completely
into work without further changes of the environment. The Szilard engine is designed as a possible realization
of such a perpetuum mobile but in the present paper we will concentrate on the entropy balance, against the grain, so to speak.

The one-molecule gas is initially confined to a cylindrical box ${\mathcal V}$ with volume ${\sf V}$ that will be separated
into two chambers ${\mathcal V}_R$ and  ${\mathcal V}_L$  with equal volumes ${\sf V}/2$ by the adiabatic insertion of a piston.
Contrary to \cite{Z84} we will neg\-lect the preparatory process of insertion of the piston since
it is only needed for a cyclic process but would be irrelevant for the entropy balance.
The Hilbert space of the gas will be chosen as
\begin{equation}\label{AS1}
  {\mathcal H}_g = {\mathcal H}_R\oplus  {\mathcal H}_L\equiv  L^2  \left({\mathcal V}_R\right)\oplus L^2  \left({\mathcal V}_L\right)
  \;.
\end{equation}

The isothermal expansion cannot be described by a unitary operator acting only on ${\mathcal H}_g$.
Thus we have to extend the object system by a heat bath with Hilbert space ${\mathcal H}_b$ and
take the Hilbert space of the object system as
\begin{equation}\label{AS2}
   {\mathcal H}={\mathcal H}_g\otimes {\mathcal H}_b
   \;.
\end{equation}
We note that these Hilbert spaces are infinite-dimensional. Strictly speaking, we are restricted
to the finite-dimensional case according to the general assumption in Section \ref{sec:OI} but
we do not expect that this will cause any problem.

Initially the state of the object system is assumed to be given by the  product state
\begin{equation}\label{AS3}
 \rho_0=\rho_g\otimes \rho_b\equiv \frac{1}{2}\left( \rho_R + \rho_L\right)\otimes \rho_b
\end{equation}
where $\rho_R$, $\rho_L$ and $\rho_b$ are Gibbs states with the same temperature $T$ corresponding
to suitable Hamiltonians. The Hamiltonian for the gas is the one-particle kinetic energy with
the boundary condition  of vanishing wave functions at the boundaries of ${\mathcal V}_R$ and ${\mathcal V}_L$.
Due to symmetry considerations we will assume
\begin{equation}\label{AS6}
  S( \rho_R)=S( \rho_L)
  \;.
\end{equation}

The projectors of the first L\"uders measurement will be $P_R$ and $P_L$ corresponding
to projections onto the subspaces  ${\mathcal H}_L$ and ${\mathcal H}_R$, resp., see (\ref{AS1}).
These projectors commute with $\rho_0$ and thus the corresponding total L\"uders operation (\ref{OI1}) alone would
not change the state $\rho_0$. But we have to perform a conditional action:
Depending on the result of this measurement one of two possible isothermal expansions will be performed that are
described by unitary operators
$U_R,\,U_L:  {\mathcal H} \rightarrow  {\mathcal H}$.
Hence the state of the object system after the conditional action will be
\begin{equation}\label{AS8}
 \rho_1=\frac{1}{2}\left(U_R\left( \rho_R\otimes \rho_b\right) U_R^\ast + U_L \left(\rho_L\otimes \rho_b\right) U_L^\ast\right)
 \;.
\end{equation}
One expects from physical reasons that after the isothermal expansion
one would obtain a one-dimensional gas filling the box ${\mathcal V}$ in thermal equilibrium
with the heat bath. Hence
both density operators in (\ref{AS8}) will be equal to a Gibbs state of the
form $\rho_g\otimes \rho_b$, but with a slightly lower temperature than $T$.
However, we will not need this strong
thermalization assumption but only the weaker one that can be justified by symmetry considerations:
\begin{equation}\label{AS9}
  U_R\left( \rho_R\otimes \rho_b\right) U_R^\ast
\approx
   U_L \left(\rho_L\otimes \rho_b\right) U_L^\ast
 \approx \rho_1
  \;,
\end{equation}
where the second approximation follows from (\ref{AS8}).
Eq.~ (\ref{AS9}) implies
\begin{eqnarray}
\label{AS10a}
  S(\rho_1) &\approx& S\left( U_R\left( \rho_R\otimes \rho_b\right) U_R^\ast\right) \\
  \label{AS10b}
   &=& S\left( \rho_R\otimes \rho_b\right) \\
   \label{AS10c}
   &=& S(\rho_R)+S(\rho_b)
   \;.
\end{eqnarray}

This further gives the following result for the entropy decrease due to the conditional action:
\begin{eqnarray}\nonumber
  S(\rho_1)-S(\rho_0)&\stackrel{(\ref{AS10c},\ref{AS3})}{=} & \left(S(\rho_R)+S(\rho_b)\right)- \left(S(\rho_g)+S(\rho_b) \right)\\
 && \label{AS13a}\\
 \label{AS13b}
  &=& S(\rho_R)-S(\rho_g)\\
   \label{AS13c}
  &\approx& -\log 2
  \;.
\end{eqnarray}
The last approximation (\ref{AS13c}) follows from
\begin{eqnarray}\label{AS14a}
  S(\rho_g)&\stackrel{(\ref{AS3})}{=}&S\left(\frac{1}{2} \rho_R + \frac{1}{2} \rho_L\right)\\
  \nonumber
  &\stackrel{(\ref{P6})}{=}&\frac{1}{2}\, S(\rho_R )+\frac{1}{2}\, S(\rho_L)\\
  \label{AS14b}
  && -\frac{1}{2}\,\log \frac{1}{2} -\frac{1}{2}\,\log\frac{1}{2}\\
  \label{AS14c}
  &\stackrel{(\ref{AS6})}{=}& S(\rho_R )+\log 2
  \;.
\end{eqnarray}

This entropy decrease has not been calculated directly by Zurek in \cite{Z84} but follows
from his result
\begin{equation}\label{AS15}
  \Delta A= k_B \,T \log 2
\end{equation}
for the increase of free energy $A$ of the gas due to the position measurement,
 see \cite{Z84} Eq.~(20),
if we take into account the thermodynamical identity
\begin{equation}\label{AS16}
 S\,T= E-A
 \;,
\end{equation}
and that the intrinsic energy $E$ of the gas does not change due to the measurement.
(Note that we have used dimensionless entropy units in this paper and hence set Boltzmann's constant $k_B$ to $1$.)

Zurek also considers in \cite{Z84}, section ``Measurement by `quantum Maxwell's demon' ", a measurement dilation
similar to that considered in this paper, but only for the pure L\"uders measurement, not for the conditional action.
Nevertheless, he obtained an entropy increase of $\Delta S=k_B\,\log 2$ of the demon's state that exactly compensates
the entropy decrease of the gas calculated above, and related this entropy increase to the loss of ``mutual information",
see \cite{Z84} Eq.~(36). It will be instructive to compare these considerations with the measurement dilation scheme
considered in Section \ref{sec:QMD} applied to the Szilard engine model.

We choose the demon's Hilbert space as ${\mathcal K}={\mathbbm C}^2$ with orthonormal basis $(r,\ell)$ and
projectors $Q_R=|r\rangle \langle r|,\; Q_L=|\ell\rangle \langle \ell|$. The initial state of the demon
will be chosen as $\phi=r$.
Further we choose a unitary operator
$V: {\mathcal H}_g \otimes {\mathcal H}_b \otimes {\mathcal K}  \rightarrow  {\mathcal H}_g \otimes {\mathcal H}_b \otimes {\mathcal K}$
satisfying
\begin{eqnarray}
\label{AS17a}
 V\left( \psi_R\otimes \psi_b\otimes r\right) &=& U_R\left(\psi_R\otimes \psi_b\right)\otimes r,  \\
 \label{AS17b}
 V\left( \psi_L\otimes \psi_b\otimes r\right) &=& U_L\left(\psi_L\otimes \psi_b\right)\otimes \ell
 \;,
\end{eqnarray}
for all $\psi_R\in{\mathcal H}_R,\;\psi_L\in{\mathcal H}_L,$ and $\psi_\in {\mathcal H}_b$.

The factors of the initial state $\rho_0=\rho_g\otimes \rho_b$
will have  spectral decompositions of the following form
\begin{eqnarray}
\nonumber
 \rho_g&=&\frac{1}{2}\left( \rho_R + \rho_L\right) =\frac{1}{2} \sum_i p_i\left(
 |\psi_R^{(i)}\rangle \langle  \psi_R^{(i)}|+|\psi_L^{(i)}\rangle \langle  \psi_L^{(i)}|
  \right)\\
  &&\label{AS18a}\\
  \label{AS18b}
  \rho_b &=& \sum_j b_j|\psi_b^{(j)}\rangle \langle  \psi_b^{(j)}|
  \;,
\end{eqnarray}
where we have used that the eigenvalues $p_i$ of $\rho_R$ are the same as those for $\rho_L$ due to symmetry.

After a straight forward calculation using (\ref{AS18a}) and (\ref{AS18b})
we obtain for the total state $\rho_{12}$ after the interaction the expected result
\begin{equation}\label{AS19}
\rho_{12}=V\left(\rho_g\otimes \rho_b\otimes |r\rangle \langle r|\right) V^\ast =
\rho_1\otimes \frac{1}{2}\left( |r\rangle \langle r|+|\ell\rangle \langle \ell|\right)
\;,
\end{equation}
with $\rho_1$ according to (\ref{AS9}). Since $\rho_{12}$ commutes with ${\mathbbm 1}\otimes Q_R$ and ${\mathbbm 1}\otimes Q_L$
the final L\"uders measurement does not change this state:
\begin{equation}\label{AS20}
 \rho_{12}= \left({\mathbbm 1}\otimes Q_R \right) \rho_{12} \left({\mathbbm 1}\otimes Q_R \right)+
 \left({\mathbbm 1}\otimes Q_L \right) \rho_{12} \left({\mathbbm 1}\otimes Q_L \right)
 \;,
\end{equation}
analogously to the measurement dilation considered in \cite{Z84}. The difference to our calculation is
that we have no correlation between object system and demon in the final state $\rho_{12}$ and the
separation into partial traces considered in Section \ref{sec:OI} is superfluous.

Consequently, the total entropy during the conditional action will be constant since the entropy of the demon increases by $\Delta S_d= \log 2$
as can be directly read off the final demon's state in (\ref{AS19}), and the entropy of the object system decreases
according to $S(\rho_0)-S(\rho_1)=-\log 2$, see (\ref{AS13b}) and (\ref{AS14c}).

We should add a remark on the role of approximations in the present problem of Szilard's engine.
These approximations simplify the presentation but are not crucial for the total entropy balance that is guaranteed by
the measurement dilation as explained in Section \ref{sec:QMD}. For example, if we cancel the approximation,
that the isothermal expansion reaches the same state in both cases, see (\ref{AS9}),
then in the final state $\rho_{12}$ after the interaction a small correlation would remain.
The following measurement and the separation of the states of subsystems would lead to a small
further increase of entropy without changing the final result substantially.
A variant of the Szilard engine without any need for approximations would be
obtained by replacing the final isothermal expansion by an adiabatic expansion without any heat bath.
Of course this runs counter to the original motive of constructing a cyclic heat engine.

%%%%%%%%%%%%%%%%%%%%%%%%%%%%%%%%%%%%%%%%%%%%%%%%%%%%%%%%%%%%%%%%%%%%%%%%%%%%%%%%%%%%%%%%%%%%%%%%%%%%%%%%%%%%%%%%%%%%%%%%%%%%%%%%%%%%%%%%%%%%%%%%%
\begin{acknowledgments}
I thank all members of the DFG Research Unit FOR 2692
as well as Thomas Br\"ocker for stimulating and
insightful discussions and hints to relevant literature.
\end{acknowledgments}
%%%%%%%%%%%%%%%%%%%%%%%%%%%%%%%%%%%%%%%%%%%%%%%%%%%%%%%%%%%%%%%%%%%%%%%%%%%%%%%%%%%%%%%%%%%%%%%%%%%%%%%%%%%%%%%%%%%%%%%%%%%%%%%%%%%%%%%%%%%%%%%%%

%\newpage

\end{document}